\documentstyle[amssymb,12pt]{article}
\begin{document}
\begin{titlepage}
\thispagestyle{empty}
\begin{flushright}
BROWN-HET-1185\\
KIAS-P99040
\end{flushright}
\vspace{0.5in}
\begin{center}
{\Large \bf Lepton flavor mixing and Baryogenesis
\\}
\vspace{0.7in}
{\bf Kyungsik Kang$^1$, Sin Kyu Kang$^2$ and Utpal Sarkar$^3$\\}
\bigskip
{$^1$ \sl Department of Physics, Brown University, Providence, 
RI 02912, USA\\}
\bigskip
{$^2$ School of Physics, Korea Institute for Advanced Study, Seoul 130-012,
Korea \\}
\bigskip
{$^3$ \sl Physical Research Laboratory, Ahmedabad 380 009, India \\}
\bigskip
\end{center}

\begin{abstract}

Recently a new general class of quark mass matrix ansatz has been
proposed, which originates from some flavor symmetry. We extend 
that symmetry to the lepton sector and study the neutrino mass matrix
and address the question of the baryon asymmetry of the universe in this
model.

\end{abstract}
\end{titlepage}

The question of flavor mixing and fermion masses can lead us to 
physics beyond the standard model, where several experimental results
are present without any theoretical insight.
As an attempt to derive relationship between the quark masses and flavor
mixing hierarchies, mass matrix ansatz was suggested about two decades ago
\cite{mat}
%Among many attempts are
%some phenomenological consideration, where an ansatz of the quark
%mass matrix is considered with the hope to calculate the mixing 
%matrix \cite{mat}. 
These ansatz hopefully will emerge from some 
definite theoretical consideration.

Of the several ansatze of quark mass matrices, the canonical mass matrices
of the Fritzsch-type have been generally taken to predict
the entire Cabbibo-Kobayashi-Maskawa (CKM) matrix \cite{frit}. However, this 
ansatz is ruled out because it predicts a top quark mass to be no longer
than 100 GeV \cite{top}.
Recently this form of mass matrices has been generalized to accommodate
the large top quark mass while keeping the {\it calculability} of 
the ansatz \cite{kk}.
It was shown that this generalized mass matrix could originate from 
breaking of the maximal permutation symmetry \cite{perm1,perm2}. 

In this letter we study the consequence of the breaking of the maximal
permutation symmetry in the lepton sector. Since this symmetry acts
on both the left- and the right-handed quark fields, 
when we generalize it to the lepton sector,  
it would be natural to include the right-handed neutrinos. 
Then the same symmetry could act on the left- and right-handed leptonic 
fields in the same way as in the quark sector. 
We shall assume that lepton number is broken at some intermediate scale
when the right handed neutrinos get a Majorana mass. Then the usual
Higgs doublets will combine the left-handed neutrinos to the right-handed
neutrinos through the Dirac mass term, so that the left-handed neutrinos 
get a small see-saw Majorana mass \cite{seesaw}. There is no 
$SU(2)_L$ triplet Higgs scalar \cite{ma}, which can break lepton number 
at some high scale and give a Majorana mass to the left-handed neutrinos.
But, the maximal permutation symmetry will determine the form of the 
Majorana mass matrix for the right handed neutrinos and the Dirac neutrino mass
matrix. From these matrices the see-saw mass matrix of
the left-handed neutrinos will also be determined. 
With these neutrino mass matrices of the 
left-handed and the right-handed Majorana particles we shall study
the possibility of generating a lepton asymmetry of the universe 
which will get converted to a baryon asymmetry of the universe
before the electroweak phase transition \cite{bary1,fy1,ht}. 

Let us begin by introducing the general form of the Fritzsch-type  mass
matrix in which a nonvanishing (2,2) element
is introduced to accommodate a large top quark mass\cite{kk}.
This mass matrix form is achieved by breaking the democratic flavor symmetry
$S(3)_L \times S(3)_R \to S(2)_L \times S(2)_R \to S(1)_L \times S(1)_R$.
The resulting quark mass matrix takes the form,
\begin{equation} 
M_q = \pmatrix{ 0&A&0 \cr A&D&B \cr 0&B&C }.
\end{equation}
Although this matrix contains four independent parameters even in the case of
real parameters, one can make additional ansatz to relate $D$ to $B$ in general
so as to maintain the {\it calculability} \cite{kk}.

We shall now assume that the same flavor symmetry is also true in the lepton
sector.  In the $S(3)_L \times S(3)_R $ symmetric limit the charged lepton 
mass matrix has the form in the democratic basis:
\begin{eqnarray}
\frac{a}{3}\left( \begin{array}{ccc}
                  1 & 1 & 1 \\
                  1 & 1 & 1 \\
		  1 & 1 & 1   \end{array} \right),
\end{eqnarray}
which will lead to the largest contribution in the mass eigenstate basis.
By making unitary transformation of the mass matrix Eq.(2) with the help of
\begin{eqnarray}
U=\left( \begin{array}{ccc}
  \frac{1}{\sqrt{2}} & -\frac{1}{\sqrt{2}} & 0 \\
  \frac{1}{\sqrt{6}} & \frac{1}{\sqrt{6}} & -\frac{2}{\sqrt{6}} \\
  \frac{1}{\sqrt{3}} & \frac{1}{\sqrt{3}} & \frac{1}{\sqrt{3}} 
               \end{array} \right),
\end{eqnarray}
one can obtain the diagonal mass matrix with only a nonvanishing (3,3) element 
$m_{33}=a$. 
In order to account for the second generation masses, 
we break the $S(3)_L \times S(3)_R$ symmetry 
to $S(2)_L \times S(2)_R$.
This can be achieved by adding the following matrix to Eq.(2):
\begin{equation}
\pmatrix{ 0 & 0 & b \cr 0 & 0 & b \cr b & b & c}.
\end{equation}
%
%The $S(2)_L \times S(2)_R$
%symmetry makes non-zero elements of the mass matrix to be
%$m_{13} = m_{23} = m_{31} =m_{32} =b$ and $m_{33} =c$. 
Although different choice of the parameters $b$ and
$c$ is  general, 
in this paper we set for simplicity, $b=-c$, which renders
the mass matrix diagonal after unitary transformation with Eq.(3). 
We then get the second largest mass given by
$m_{22}=2b$. 
Finally, the $S(1)_L \times S(1)_R $ symmetric mass matrix can
be achieved by taking (1,2) and (2,1) elements to be nonzero
$(= d )$, which can then be rotated to the diagonal form with
$m_{11}=d$, $m_{22}=2b + d$ and $m_{33} = a + c$. 

We shall now work in the basis in which the charged lepton mass matrix
is diagonal and real. For the $S(3)_L \times S(3)_R$ symmetry to be applicable
we further consider that the model contains three right-handed neutrinos
$N_\alpha, \alpha=1,2,3$ in addition to the usual left-handed neutrinos of 
three generations. 
Furthermore, since the right-handed neutrinos are singlets
under the standard model gauge group they can get 
Majorana masses at the tree level, 
whereas the left-handed neutrinos can not get any tree level Majorana
mass since there are no $SU(2)_L$ triplet Higgs scalar. 
The left-handed neutrinos then remain light and get a see-saw 
mass \cite{seesaw} due to their usual Dirac type coupling 
with the right-handed neutrinos.
We can then write the neutrino mass matrix in the basis 
$[\nu_{iL} ~~ N_{\alpha R}]$ as,
\begin{equation} 
{\cal M}_\nu=\pmatrix{ 0&m_D \cr m_D^T & M_N}
\end{equation}
where each of these elements are $3 \times 3$ matrices. 

Assuming that the flavor symmetry under consideration comes from a
transformation of the left- and the right-handed fields, the right-handed
Majorana mass matrix will have a $S(3)_R \times S(3)_R$ symmetry to
start with, which will then break to $S(2)_R \times S(2)_R$ and then to
$S(1)_R \times S(1)_R$ in succession. The $S(3)_R \times S(3)_R$ symmetry
gives a mass matrix of the form \cite{kk2},
\begin{equation}
M_N = \pmatrix{ A &B &B \cr B& A& B \cr B &B&A}
\end{equation}
in which we can choose $B=0$ without loss of generality. Otherwise, we 
can keep the $B$ and diagonalize it with the unitary matrix
with the first row and the third row interchanged from Eq.(2)
to a form proportional to a unit matrix $M_N = A $ {\bf I}. 
Then the combination $N_1+N_2+N_3$ gets
a mass, which we identify with the largest mass of the right-handed neutrino
$N_{e R}$. Identification of the right-handed electron neutrino as
the heaviest one is neccessary to explain the large mixing between
the left-handed $\mu$ and $\tau$ neutrinos, which is required to
explain the atmospheric neutrino anomaly. We shall discuss this
point in details at some later stage.
In the same way, the $S(2)_R \times S(2)_R$ 
symmetry will contribute to the (2,2) diagonal elements by an amount
$C$, and the corresponding state may be identified as the $N_{\mu R}$.
Finally, the $S(1)_R \times S(1)_R$ symmetry will contribute only to the 
(3,3) element by an amount $D$. The resulting matrix in the basis
$[N_{e R} ~~ N_{\mu R} ~~ N_{\tau R}]$ then becomes,
\begin{equation}
M_N = {\rm diag}[A ,~~~ C, ~~~ D].
\end{equation}
We now assume that as in the quark sector the largest contribution to
the left-handed neutrino mass comes
in the $S(3)_R \times S(3)_R$ symmetric limit, then the heavy right-handed
neutrinos follow an inverted hierarchical pattern $M_{Ne} \gg M_{N\mu } 
\gg M_{N\tau }$ and $N_{\tau R}$ becomes the lightest right-handed
Majorana neutrino. As we shall demonstrate, this inverted hierarchical
pattern can explain the present
experimental results of the neutrino masses (which is given in table 1). 

\begin{table}
\caption{Present experimental constraints on neutrino masses and mixing}
\begin{center}
\begin{tabular}{||rcl||}
\hline \hline &&\\
Solar Neutrino {\cite{sol}}
&:& $\Delta m^2 \sim (0.5 - 1) \times 10^{-5} eV^2$ \\ 
(Small angle MSW) \phantom{[1]}
&& $ \sin^2 2 \theta \sim 10^{-2} - 10^{-3}$ \\ &&\\
Atmospheric Neutrino \cite{atm}
&:&$ \Delta m^2 \sim (0.5 - 6) \times 10^{-3} eV^2$ \\
&&$ \sin^2 2 \theta > 0.82 $\\ &&\\
Neutrinoless  Double  Beta  Decay \cite{bbn} &:&
$m_{\nu_e} < 0.46 eV$ \\ &&\\
CHOOZ \cite{chooz} &:& $\Delta m_{e X}^2 < 10^{-3} eV^2 $\\
&&(or $\sin^2 2 \theta_{eX} < 0.2)$ \\ &&\\
\hline \hline
\end{tabular}
\end{center}
\end{table}

As it is well known, 
it is not possible to explain the atmospheric neutrino anomaly \cite{atm},
solar neutrino problem \cite{sol} and the LSND result \cite{lsnd}
simultaneously in a three generation model,
since the mass squared differences required in the three cases are widely
different. One needs at least one more light sterile neutrino state which
mixes with the ordinary neutrinos. On the other hand, such sterile neutrinos 
are severely constrained by the present limit on the nucleosynthesis bounds
\cite{ht},
which can hardly accommodate a fourth neutrino. So, in our analysis we 
consider only three light neutrinos and thus do not try to accommodate the 
LSND result. There are three allowed regions of the parameter space for 
explaining the solar neutrino data \cite{soldata}, 
out of which we consider only the
small angle MSW solution, which comes out naturally in this model.

The structure of the Dirac mass matrix in the neutrino sector will be 
similar to that of the quark sector and is given by,
\begin{equation}
m_D = \pmatrix{ 0 & x & 0 \cr x & t & y \cr 0 & y & z}  .
\end{equation}
However, the hierarchy in the different elements could, in general, be 
different. So, we try to determine the different parameters from experimental
inputs, rather than justifying them from some theoretical reasonings.

At this point we can justify the requirement 
for the inverted hierarchy of the right-handed Majorana neutrino masses.
Thanks to the largely hierarchical right-handed Majorana neutrino mass 
matrix, the main contribution to the left-handed Majorana masses
comes from the lightest right-handed neutrinos in the see-saw mechanism.
If $N_{eR}$ becomes the lightest right-handed Majorana neutrino,
then the largest element of the effective left-handed Majorana mass
will be the $[\mu \mu]$ element only. On the other hand, the 
atmospheric neutrino anomaly requires a large mixing between
$\nu_\mu$ and $\nu_\tau$, which requires the $[\mu \tau]$ and 
$[\tau \mu ]$ elements to be comparable or larger than the $[\mu \mu]$
element. Thus this cannot explain the large $\nu_\mu$ and $\nu_\tau$
mixing. If we assume that $N_{\tau R}$ is the lightest right-handed 
Majorana neutrino and $y$ and $z$ are of the same order of magnitude
\cite{dp},
then all the four elements $[\mu \mu], [\mu \tau], [\tau \mu]$ and 
$[\tau \tau]$ will be of the same order of magnitude, which ensures
maximal mixing between $\nu_\mu$ and $\nu_\tau$ \cite{bar}. 

Along with the inverted hierarchy condition, $M_{N_e} \gg M_{N_\mu}
\gg M_{N_\tau}$, we also assume that the elements of the Dirac mass 
matrix are not largely hierarchical. Then the largest contribution to the 
effective $3 \times 3$ mass matrix of the left-handed neutrinos will 
come from the see-saw contribution from $ M_{N_\tau}$ and then 
the next leading contribution will come from $ M_{N_\mu}$. 
Then the effective $3 \times 3$ neutrino mass matrix can now 
be written as
\begin{eqnarray}
m_{\nu} &=& m^{T}_D M^{-1}_N m_D \nonumber \\
      &\simeq & \pmatrix{
         \frac{x^2}{M_{N_\mu}} & \frac{xt}{M_{N_\mu}} & \frac{xy}{M_{N_\mu}} \cr
          \frac{xt}{M_{N_\mu}} & \frac{y^2}{M_{N_\tau}} & \frac{yz}{M_{N_\tau}} 
	  \cr
          \frac{xy}{M_{N_\mu}} & \frac{yz}{M_{N_\tau}} & \frac{z^2}{M_{N_\tau}}}
\end{eqnarray}
where we have ignored the next to leading terms of order $O(1/M_{N_e})$
and $O(1/M_{N_\mu}^2)$. This  means that the mass squared difference 
between $\nu_\mu$ and $\nu_\tau$ vanishes. So, for the 
nonvanishing mass squared difference we need to keep the higher
order terms. However, to calculate the mixing matrix, we can ignore
the next to leading terms without loss of generality. 

Since we are interested in the maximal mixing solution
for $\nu_{\mu}$ and $\nu_{\tau}$
and the small mixing solution for $\nu_e$ and $\nu_{\mu}$,
the neutrino mixing matrix can be parameterized by
\begin{eqnarray}
U_{\nu} &=& 
\pmatrix{ 1 & 0 & 0 \cr 0 & c_2 & s_2 \cr 0 & -s_2 & c_2} \cdot
\pmatrix{ c_1 & s_1 & 0 \cr -s_1 & c_1 & 0 \cr 0 & 0 & 1}  \nonumber \\
%  \left(\begin{array}{ccc}
%    1 & 0 & 0 \\
%    0 & c_2 & s_2 \\
%    0 & -s_2 & c_2 \end{array}\right)
% \left(\begin{array}{ccc}
%    c_1 & s_1 & 0 \\
%    -s_1 & c_1 & 0 \\
%    0 & 0 & 1     \end{array}\right) \\
    &=&  
\pmatrix{ c_1 & s_1 & 0 \cr
   -\frac{s_1}{\sqrt{2}} &\frac{c_1}{\sqrt{2}} & \frac{1}{\sqrt{2}} \cr
    \frac{s_1}{\sqrt{2}} &-\frac{c_1}{\sqrt{2}} & \frac{1}{\sqrt{2}}},
% \left(\begin{array}{ccc}
%    c_1 & s_1 & 0 \\
%   -\frac{s_1}{\sqrt{2}} &frac{c_1}{\sqrt{2}} & \frac{1}{\sqrt{2}} \\
%    \frac{s_1}{\sqrt{2}} &-frac{c_1}{\sqrt{2}} & \frac{1}{\sqrt{2}} \\
%      \end{array}\right)
\end{eqnarray}
where $c_1=\cos\theta_1, c_2=\sin \theta_2$ and we have taken 
$c_2=s_2=1/\sqrt{2}$.
Then, the favored solution for the solar and atmospheric neutrino anomaly
leads to $c_1\sim 1, s_1 \sim 0.05, m_2\sim 3\times 10^{-3}$ eV
and $m_3\sim 3\times 10^{-2}$ eV.
{}From these results, one can approximately estimate the neutrino mass matrix
\begin{eqnarray}
m_{\nu}&=& U_{\nu}\cdot diag[m_1,-m_2,m_3]\cdot U_{\nu}^{\dagger} \\
&\sim& \pmatrix{ 10^{-6} & 10^{-4} & 10^{-4} \cr
 10^{-4} & 1.5\times 10^{-2} &   1.5\times 10^{-2}\cr
  10^{-4} & 1.5\times 10^{-2} &   1.5\times 10^{-2}}
% \left( \begin{array}{ccc}
%10^{-6} & 10^{-4} & 10^{-4} \\
% 10^{-4} & 1.5\times 10^{-2} &   1.5\times 10^{-2}\\
%  10^{-4} & 1.5\times 10^{-2} &   1.5\times 10^{-2}
%  \end{array}\right)
\end{eqnarray}
%{\bf Comments : to include more details of the mixing matrix etc, which
%gives the above results. In baryogenesis some more expressions has to be
%presented.}
In this case, by choosing $y$ and $z$ to be of the same order of magnitude
we can ensure the maximal mixing between $\nu_\mu$ and $\nu_\tau$. 
To be precise, the large mixing between $\nu_\mu$ and $\nu_\tau$, as 
observed in the atmospheric neutrinos at SuperKamiokande(i.e., $\sin^2 2\theta
>0.82$), implies $0.64 < y/z < 1.56$. 
On the other hand, the ratio $x/y \sim x/t$ gives the
mixing between $\nu_e$ and $\nu_\mu$ and could be very small. 
For a reasonable choice, $x/y \sim x/t \sim 10^{-2}$, the solar 
neutrino anomaly can be explained by the small angle  MSW solution, which is one
of the favored solutions for the solar neutrino anomaly \cite{soldata}.

In all the models for neutrino masses, another related question remains
to be answered is the generation of the baryon asymmetry of the universe.
The Majorana mass term of the right handed neutrinos introduces 
lepton number violation  and hence $(B-L)$ violation. If this interaction
is slower than the expansion rate of the universe when the right handed
neutrinos decay (at $T = M_N$) and there is enough $CP$ violation, 
this interaction can generate a lepton asymmetry of the universe. 
At finite temperature above the critical temperature of the electroweak
phase transition sphaleron processes are in thermal equilibrium and
(B$+$L) number violating interactions are very fast \cite{krs}. This
will then relate the lepton asymmetry or the $(B-L)$ asymmetry 
generated during the right handed neutrino decay to the baryon 
asymmetry of the universe before the electroweak phase transition. 
This remains to be the most interesting scenario for the 
understanding of the baryon number of the universe,
which is referred to as leptogenesis \cite{fy1,lg}.

Since lepton number violation is the source of leptogenesis, it also
depends on models of neutrino masses.
It has recently been argued \cite{dav} that the see-saw mechanism 
of neutrino masses is the most
preferred one for the generation of the baryon asymmetry of the 
universe in supersymmetric inflationary models.
We shall now see if this could be implemented in the present scenario. 

In the see-saw mechanism for neutrino masses the right-handed 
neutrino decay could generate a lepton asymmetry of the universe, which
then gets converted to a baryon asymmetry of the universe. CP violation
comes from an interference of the tree level diagrams with the self-energy
type and vertex correction type diagrams \cite{fy1,lg}. 
We start with the lagrangian
\begin{equation}
{\cal L} = M_i \overline{N^c_{i R} } N_{i R} + h_{\alpha i} 
\overline{\ell_{\alpha L }} \phi N_{i R} + h.c.,
\end{equation}
where $\ell_{\alpha L}$ are the light leptons, $\phi$ is the usual
Higgs doublet of the standard model, $h_{\alpha i}$ are the complex
Yukawa couplings and $\alpha$ is the generation index.
We have chosen the Majorana mass matrix of the right-handed 
neutrino to be real and diagonal.
Then the decay of the lightest neutrino $N_{\tau R}$ can generate a lepton
asymmetry which can then generate a baryon asymmetry of the universe 
before the electroweak phase transition. The decay of the heavier neutrinos
can also generate an asymmetry, but that asymmetry will be washed out
before the lightest right-handed neutrino decay. 

Due to the Majorana masses of the right-handed neutrinos, their decay
violates lepton number,
\begin{eqnarray}
N_{i R} &\longrightarrow & \ell_{\alpha L} + \phi^{\dagger} \\
& \longrightarrow & \ell^c_{\alpha L} + \phi .
\end{eqnarray}
$CP$ violation comes from an interference of the tree level diagram for 
the decay of $N_{i R}$ with the one loop diagrams shown in figure 1.
There are two contributions coming from the interference of the tree 
diagram with the one loop vertex correction and another one with the one 
loop self energy diagram. When the masses of the heavy neutrinos are 
degenerate, there can be large contributions coming from an 
interference of the self energy diagrams. However, in the present case
when the masses of the right handed neutrinos are hierarchical, the
contribution coming from the two sources are equal. In this case the
contributions from both the diagrams add up to give the total amount
of $CP$ violation. First, the heaviest right handed neutrino will
decay and then the second heaviest, when it may or may not generate
any lepton asymmetry. When the lightest one ($N_{\tau R}$) decays,
it will first erase any pre-existing asymmetry and then generate the
final lepton asymmetry, which is presented by
\begin{eqnarray}
\epsilon_\tau &=&\frac{\sum_\alpha \Gamma(N_{\tau R}\rightarrow  
\ell_{\alpha L}\phi^{\dagger})
-\sum_\alpha \Gamma(N_{\tau R}\rightarrow \ell^c_{\alpha L}\phi)}
{\sum_\alpha \Gamma(N_{\tau R}\rightarrow  \ell_{\alpha L}\phi^{\dagger})
+\sum_\alpha \Gamma(N_{\tau R}\rightarrow \ell^c_{\alpha L}\phi)} \nonumber \\
&& \nonumber \\
&=& {3 \over 16 \pi} {{\rm Im} [ \sum_\alpha 
(h^\ast_{\alpha \tau} h_{\alpha i}) \sum_\beta 
(h^\ast_{\beta \tau}  h_{\beta i})] \over \sum_\alpha 
|h_{\alpha \tau} h^\ast_{\alpha \tau}|^2} I({M_{N_\tau}^2 \over 
M_{N_i}^2}) ,
\end{eqnarray}
where $i=e,\mu$ and $I(x)=x^{1/2}[1+(1+x)\ln (x/(1+x))]$.  \\

Similar to the Jarlskog invariant for $CP$ violation in the quark sector,
in this case a different combination enters, which has its origin in 
the Majorana nature of the neutrino masses. The Majorana nature of the
right handed neutrinos will imply that there are new Majorana phases
which will contribute to the $CP$ violation. In the basis we are working,
where the right handed mass matrix is real and diagonal, all these
Majorana phases has been transferred to the elements of $m_D$ and
the quantity $\sum_\alpha (h^\ast_{\alpha \tau} h_{\alpha i})$
becomes a rephasing invariant quantity. In the limit $x = 0$, this
becomes equivalent to a two heavy neutrino scenario. In this case
after considering the overall rephasing there remains only one $CP$ 
phase, which vanishes when $t=0$. Thus, in the Fritzsch type of 
mass matrix, the $CP$ violation for the generation of the a baryon
asymmetry of the universe will be highly suppressed.

In terms of the elements of $m_D$, the amount of $CP$ violation is
given by,
\begin{equation}
\epsilon_\tau = {3 \over 16 \pi} \left[ {{\rm Im}(y^\ast t + z^\ast y)^2
\over (|y|^2 + |z|^2)}\left( {M_{N_\tau} \over M_{N_\mu}} \right)
+ {{\rm Im}(y^\ast x)^2
\over (|y|^2 + |z|^2)}\left( {M_{N_\tau} \over M_{N_e}} \right) \right].
\end{equation}
We can now choose the overall phase so that $z^\ast y$ is real. 
Then for $t=0$, the first term vanishes and assuming that the $CP$ 
phases are all similar the amount of CP violation gets suppressed
by an amount, $\displaystyle{x M_{N_e} \over y M_{N_\tau}} 
< 10^{-4}$. With this additional
suppression it will be impossible to explain the baryon asymmetry of
the universe. Thus, for the generation of the lepton asymmetry of
the universe, it is extremely important that we make the [2,2] 
element non-vanishing, which 
required to accommodate the heavy top quark mass in Fritzsch type
(mass matrix) ansatz.

For the generation of a lepton asymmetry of the universe one more
ingredient is neccessary, namely, the out-of-equilibrium condition.
Whether a system is in equilibrium or not can be understood by solving
the Boltzmann equations.
But a crude way to put the out-of-equilibrium condition is to say
that the universe expands faster than the interaction rate \cite{fry}.
This may be stated as
\begin{equation}
\Gamma_N < H=1.7\sqrt{g_{\star}}\frac{T^2}{M_P}
\end{equation}
where $\Gamma_N$ is the interaction rate under discussion, 
$g_{\star} \sim 10^2$ is
the effective number of degrees of freedom available at that temperature
$T$, and $M_P$ is the Planck scale. 

In the case of right-handed neutrino decay, the asymmetry is generated
when the lightest right handed neutrino decays. Before its decay, 
the pre-existing lepton asymmetry, if any, is washed out by its lepton 
number violating interactions. 
Just before the lightest right handed neutrino decays, it satisfies
the out-of-equilibrium condition 
\begin{equation}
\frac{|h_{\alpha 1}|^2}{16\pi}M_N < H ~~~\mbox{at}~~T=M_N
\end{equation}
if the masses of the
heavy neutrinos are larger than $M_N > 10^7 $ GeV.  
We consider that the reheating temperature is not too high so
that after reheating gravitinos are not produced in large numbers which
can then overclose the universe \cite{buch}. This implies that for the 
lightest right-handed neutrinos to be produced after reheating, we 
must have $M_{N \tau } < 10^{10}$ GeV. 
This reduces the uncertainty in the scale 
of the heavy neutrinos to some extent. The explanation of the atmospheric 
neutrino anomaly, i.e., the mass of $\nu _\tau$ to be around $10^{-2}$ 
then requires, $y,z,t \sim 0.1 $ GeV for $M_{N \tau } \sim 10^8$ GeV. 
Taking the vacuum expectation value
of the Higgs doublet field to be around 100 GeV, we get a lepton 
asymmetry of the universe to be around, $n_L \sim {\epsilon \over g_*}
\sim 10^{-8} \sin \delta $, where $\delta $ is the 
$CP$ phase in the couplings of the Dirac neutrino mass
matrix $m_D$. So, an $CP$ violating phase of the order
of $\sin \delta \sim 0.01$ can generate enough lepton asymmetry of 
the universe. During the period $10^{12}~~ {\rm GeV} > T >
10^2~~ {\rm GeV}$ the sphaleron transitions will be very fast and
this lepton asymmetry
will get converted to a baryon asymmetry of the universe, $n_b \sim {n_L 
\over s}$.

In summary, we extended a new general class of quark mass matrix ansatz
to the leptonic sector 
to obtain the neutrino mass matrix. We showed that an 
interesting inverted hierarchical pattern for the heavy right-handed
neutrinos can accommodate the atmospheric neutrino oscillation with the maximal 
$\nu_\mu \to \nu_\tau$ mixing and the small angle MSW solution of the
solar neutrino deficit. It turned out that the baryon asymmetry of the universe
comes out to be correct for this particular form of neutrino mass matrices.

\vskip .2in

{\bf Acknowledgments} 
One of us (US) would like to thank the organisers of the journal club
and the Korea Institute for Advanced Study for
hospitality, where this work was initiated.

\newpage
\begin{figure}[htb]
\mbox{}
\vskip 5in\relax\noindent\hskip .6in\relax
\includegraphics{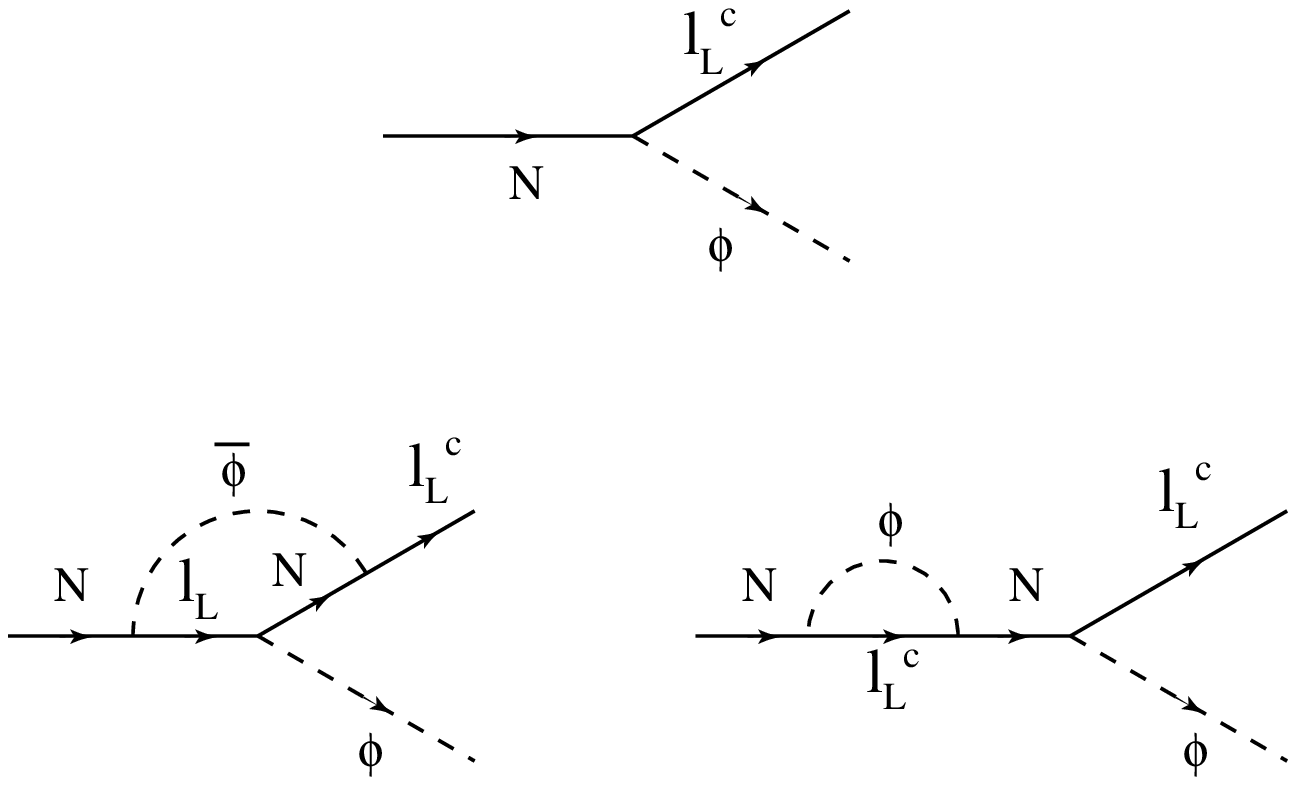}
\caption{ Tree level and one loop vertex and self energy diagrams for
the generation of lepton asymmetry of the universe }
\end{figure}


\newpage
\begin{thebibliography}{99}
\bibitem{mat} S. Weinberg, in a Festschrift for I. I. Rabi, ed. L. Motz
             (N.Y. Academy of Sciences, N.Y. 1977); A. deRejula, H. Georgi,
             and S. L. Glashow, Ann. Phys. (N.Y.) {\bf 109}, 258 (1977);
             F. Wilczek and A. Zee, Phys. Lett. {\bf B79}, 418 (1977);
             H. Georgi and D. V. Nanopoulos, Nucl. Phys. {\bf B155},
             52 (1979).

\bibitem{frit} A. C. Rothman and
             K. Kang, Phys. Rev. Lett. {\bf 43}, 1548 (1979); Phys. Rev.
             {\bf D24}, 167 (1981); H. Fritzsch, Phys. Lett. {\bf B73},
             317 (1978); Nucl. Phys. {\bf B155}, 189 (1979).

\bibitem{top}K. Kang and S. Hadjitheodoridis, Phys. Lett. {\bf B193},
              504 (1987); K. Kang, J. Flanz, and E. Paschos, Z. Phys.
              C {\bf 55}, 75 (1992) See also H. Harari and and Y. Nir,
              Phys. Lett. {\bf B 195}, 586 (1987).


\bibitem{kk} K. Kang and S. K. Kang, Phys. Rev. {\bf D 56}, 1511 (1997);
            hep-ph/9802330.

\bibitem{perm1} H. Fritzsch and Z. z. Xing, Phys. Lett. {\bf B 413}, 396 (1997);
              hep-ph/9904286; A. Mondragon, E. Rodriguez-Jauregui, Phys. Rev. 
	      {\bf D 59}, 093009 (1999); hep-ph/9906429; D. Falcone, 
	      hep-ph/9905316; M. Randhawa, V. Bhatnagar, P.S. Gill, M. Gupta,
	      hep-ph/9903428.

\bibitem{perm2} See also, S. L. Adler, Phys. Rev. {\bf D 59}, 015012 (1999);
              hep-ph/9711393. P. S. Gill and M. Gupta, Phys. Rev. {\bf D 56},
	      3143 (1997); K. Kang, S. K. Kang, C. S. Kim and S. M. Kim, 
	      hep-ph/9808419.

\bibitem{seesaw}  M. Gell-Mann, P. Ramond and R. Slansky, in {\it %
Supergravity}, eds. P. van Nieuwenhuizen and D. Freedman, (North-Holland,
1979) p.315; T. Yanagida, in {\it Proc. Workshop on Unified Theories and
Baryon Number in the Universe}, eds. A. Sawada and A. Sugamoto (KEK, 1979),
p.95; R.N. Mohapatra and G. Senjanovi\'{c}, Phys. Rev. Lett. {\bf 44}, 912
(1980).

\bibitem{ma}  E. Ma and U. Sarkar, Phys. Rev. Lett. {\bf 80} (1998) 5716;
U. Sarkar, Phys. Rev. {\bf D 59} (1999) 031301; 
E. Ma. M. Raidal and U. Sarkar, hep-ph/9811240; Eur. Phys. J. {\bf C8}
(1999) 301.

\bibitem{bary1} E. W. Kolb and M. S. Turner, 1989, {\it The Early Universe}
(Addison-Wesley, Reading, MA).

\bibitem{fy1}  M. Fukugita and T. Yanagida, Phys. Lett. {\bf B 174}, 45
(1986).

\bibitem{ht}  S. Yu. Khlebnikov and M.E. Shaposhnikov, Nucl. Phys. {\bf B 308%
}, 885 (1988); J.A. Harvey and M.S. Turner, Phys. Rev. {\bf D 42}, 3344
(1990).

\bibitem{kk2} K. Kang, S. K. Kang, J. E. Kim and P. Ko, Mod. Phys. Lett. 
{\bf A 12}, 553 (1997); K. Kang and S. K. Kang, hep-ph/9802328; S. K. Kang
and C. S. Kim, Phys. Rev. {\bf D 59}, 091302 (1999); H. Fritzsch and Z. z. Xing,
Phys. Lett. {\bf B 372}, 265 (1996); hep-ph/9807234 ; Phys. Lett. {\bf B 440}, 
313 (1998); ; M. Tanimoto, hep-ph/9807517; 
M. Fukugita, M. Tanimoto and T. Yanagida, Phys. Rev. {\bf D 57}, 4429 (1998).

\bibitem{sol}  Super-Kamiokande Collaboration : Y. Fukuda {\em et al}, Phys.
Rev. Lett. {\bf 81}, 1158 (1998); Talk by Y. Suzuki at Neutrino'98,
Takayama, Japan (1998).

\bibitem{atm}  Super-Kamiokande Collaboration : Y. Fukuda {\em et al}, Phys.
Rev. Lett. {\bf 81} (1998)1562 ; hep-ex/9805006; Phys. Lett. {\bf B433}
(1998) 9; T. Kajita, hep-ex/9810001.

\bibitem{bbn}Heidelberg-Moscow Collaboration : 
H.V. Klapdor-Kleingrothaus, in Proc Lepton and Baryon
 number violatoin, Trento, April 1998; M. G\"{u}nther et al, Phys. Rev.
 {\bf D 55} (1997) 54; L. Baudis et al, Phys. Lett. {\bf B 407} (1997) 219.

\bibitem{chooz} CHOOZ Collab., M. Apollonio {\it et. al}, hep-ex/9711002.

\bibitem{lsnd}C. Athanassopoulos {\it et. al}, Phys. Rev. Lett. {\bf 75}
(1995) 2650. 

\bibitem{soldata} N. Hata and P. Langacker, Phys. Rev. D{\bf 56} (1997) 6107;
              J.N. Bahcall, P.J. Krastev and A.Yu. Smirnov, hep-ph/9807216.

\bibitem{krs} V.A.   Kuzmin,   V.A.   Rubakov   and    M.E.
Shaposhnikov,   Phys. Lett. {\bf B 155} (1985) 36.

\bibitem{dp} E. Ma, D.P. Roy and U. Sarkar,
 Phys. Lett. {\bf B444} (1998) 391; E. Ma and D.P. Roy, Phys. Rev. {\bf D 59},
 (1999) 097702.

\bibitem{bar} R. Barbieri, L.J. Hall and A. Strumia, hep-ph/9808333.

\bibitem{lg}  P. Langacker, R. Peccei and T. Yanagida, Mod. Phys. Lett. {\bf %
A 1}, 541 (1986); A. Acker, H. Kikuchi, E. Ma and U. Sarkar, Phys. Rev. {\bf %
D 48}, 5006 (1993); M. Flanz, E.A. Paschos and U. Sarkar, Phys. Lett. {\bf B
345}, 248 (1995); L. Covi, E. Roulet and F. Vissani, Phys. Lett. {\bf B 384}%
, 169 (1996); M. Flanz, E.A. Paschos, U. Sarkar and J. Weiss, Phys. Lett. 
{\bf B 389}, 693 (1996); W. Buchmuller and M. Pl\"{u}macher, Phys. Lett. 
{\bf B 389}, 73 (1996); Phys. Lett. {\bf 431}, 354 (1998); 
J. Faridani, S. Lola, P.J. O'Donnell and U. Sarkar, Eur. Phys. J. {\bf C 7},
 543 (1999); A. Pilaftsis, Phys. Rev. {\bf D 56}, 5431 (1997); 
Nucl. Phys. {\bf B 504} (1997) 61.

\bibitem{dav} D. Delepine and U. Sarkar, hep-ph/9811479; E. Ma, 
M. Raidal and U. Sarkar, hep-ph/9901406.

\bibitem{fry}  J.N. Fry, K.A. Olive and M.S. Turner, Phys. Rev. Lett. {\bf 45%
} (1980) 2074; Phys. Rev. {\bf D 22} (1980) 2953; Phys. Rev. {\bf D 22}
(1980) 2977; E.W. Kolb and S. Wolfram, Nucl. Phys. {\bf B 172} (1980) 224.

\bibitem{buch} J. Ellis, G. B. Gelmini, J. L. Lopez, D. V. Nanopoulos and 
S. Sarkar, Nucl. Phys. {\bf B373} (1992) 399;
 M. Boltz, W. Buchmuller and M. Pl\"{u}macher, Phys. Lett. {\bf B443} 
 (1998) 209;
 D. J. H. Chung, E. W. Kolb and A. Riotto, hep-ph/9809453.


\end{thebibliography}
\end{document}